\documentclass{PoS}
\usepackage{bm}
\usepackage{epsfig}
\usepackage{graphicx}

\title{Exploration of the phase structure of SU(N$_{\bf c}$) lattice gauge theory with many Wilson fermions at strong coupling}

\ShortTitle{Exploration of the phase structure of $SU(N_c)$ lattice gauge theory}

\author{\speaker{Kei-ichi Nagai}%
         \thanks{Present address: Kobayashi-Maskawa Institute for the Origin of Particles and the Universe (KMI),  Nagoya University, Nagoya, 464-8602, Japan}\\
        KEK Theory Center, High Energy Accelerator Research Organization (KEK), 
Tsukuba, 305-0801, Japan\\
        E-mail: \email{nagaik@post.kek.jp, keiichi.nagai@kmi.nagoya-u.ac.jp}}

\author{Maria Georgina Carrillo-Ruiz,
Gergana Koleva and Randy Lewis \\
        Department of Physics and Astronomy, York University, Toronto, M3J 1P3, Canada\\
        E-mail: \email{georgecr@yorku.ca, koleva2g@uregina.ca, randy.lewis@yorku.ca}}

\abstract{We explore aspects of the phase structure 
of $SU(2)$ and $SU(3)$ lattice gauge theories at strong coupling 
with many flavours $N_f$ of Wilson fermions 
in the fundamental representation,
including the relevance to recent searches for a conformal window.
The pseudoscalar meson mass, the quark mass and other quantities are observed 
as functions of the hopping parameter,
and we find deviations from the expected analytic dependence, 
at least for sufficiently large $N_f$.
Implications of these effects for the phase structure
and for the existence of a (first order) bulk phase
and the Aoki phase 
are discussed
in the case of $N_f/N_c \gg 1$.
}

\FullConference{The XXVIII International Symposium on Lattice Field Theory, Lattice 2010\\
		June 14-19, 2010\\
		Villasimius, Italy}

\begin{document}

\section{Introduction}
\label{sec:intro}

In $SU(N_c)$ vector gauge theories,
perturbation theory allows 
for a non-trivial infrared fixed point (IRFP)\cite{caswell,Banks:1981nn}.
When an IRFP exists in the very strong coupling region,
investigation by lattice gauge theories is needed
for going beyond perturbation theory.
There have been a lot of recent lattice studies; see
Refs.~\cite{DelDebbio,Fleming:2008gy} for reviews.

We are motivated by Ref.~\cite{Iwasaki}
in which a systematic lattice analysis over a wide range
of the gauge coupling constant was performed.
According to those authors,
$N_f=3$ in $SU(2)$ and $N_f=7$ in $SU(3)$ 
belong to the conformal window.
Their result was based on the prediction
of the pion mass behaviour in the strong coupling limit,
given in Ref.~\cite{Aoki} as follows:
\begin{equation}
\label{eq:pion}
\cosh(m_\pi a) = 1 + \frac{(1-16\kappa^2)(1-4\kappa^2)}{8\kappa^2(1-6\kappa^2)}
\,,
\end{equation}
where $\kappa=\frac{1}{2 m_0 + 8}$ 
and $m_0$ is the bare mass parameter of Wilson fermions.
Eq.~(\ref{eq:pion}) implies
that the pion becomes massless 
at $\kappa=1/4$,
independent of $N_c$ and $N_f$.
Therefore
the authors of Ref.~\cite{Iwasaki} investigated 
at $\beta=\frac{2 N_c}{g^2}=0$ in particular,
which led to
their prediction for the critical flavour number
of the conformal window
mentioned above.

However,
with Wilson fermions
the phase structure in the strong coupling limit
is very complicated, 
as pointed out in Ref.~\cite{Aoki},
due to
the existence of a parity-flavour broken phase (Aoki phase).
In the Aoki phase $\kappa > \kappa_c = 1/4$ at $\beta=0$,
and on the phase boundary
the charged pion is massless
while the neutral pion is not\footnote{
The order parameter of the Aoki phase is 
$\langle \bar{\psi}\gamma_5 \tau_3 \psi \rangle$.}.

Therefore,
in Ref.~\cite{Nagai} and this proceedings
we check the particle spectroscopy
in the strong coupling limit
to determine whether Ref.~\cite{Iwasaki} is correct,
and 
we explore the phase structure
of the Wilson fermion system
with many flavours 
to determine whether the Aoki phase exists
in the many flavour case.

\section{Simulation details}
\label{sec:detail}

We use $N_f$ flavours of mass-degenerate Wilson fermions\cite{Nagai}.
Dynamical configurations are generated
by the standard Hybrid Monte-Carlo (HMC) algorithm
with $N_{MD} \Delta \tau =1$.
Lattice sizes are $6^2\times 12^2$,
$8^2 \times 16^2$, $12^2 \times 24^2$ and $12^3 \times 24$ though not all of
these sizes were used for every choice of $N_f$;
this range of sizes allows us to look for finite size effects (for some $N_f$).
In this exploration,
we compute observables from $50 \sim 100$ configurations
with $4 \sim 5$ intervals between trajectories
after thermalizing.
We monitor the following observables:
$m_\pi^2$, $m_\rho$,
the plaquette value($\langle \square \rangle$),
the axial Ward-Takahashi identity quark mass
($m_q^{\rm AWI} = \frac{\nabla_4 \langle \sum_{\bf x}A_4({\bf x},t)P({\bf 0},0)
\rangle }{2 \langle \sum_{\bf x}P({\bf x},t)P({\bf 0},0)\rangle}$),
the Polyakov loop,
Creutz ratios,
the chiral condensate (or the propagator norm),
the lowest eigenvalues,
and $\langle S(t) S(0) \rangle$ as a function of $t$.
Due to page constraints
in these proceedings
we focus on a few observables
most relevant to a discussion of the phase structure.

\section{$N_c=3$ case: Comparison with the data of Ref.~\cite{Iwasaki}}
\label{sec:su3}

Iwasaki et al.\cite{Iwasaki} concluded
that the $N_f=6$ case is in the confinement region
because they found
$N_{CG} > 10000$ 
in the Molecular-Dynamics (MD) evolution
for the thermalizing process
(they did not attain the thermalization)
and that the plaquette value was decreased
toward zero,
in contrast with the case of $N_f > 7$.
Thus they interpreted
these observations to be signals for
a massless pion,
in spite of no direct calculations of the pion mass
and the quark mass.
However,
as plotted in the left panel of Fig.~\ref{fig:nc3},
we found results for the pion and quark masses
as functions of $1/\kappa$ at $\beta=0$
that differ from Ref.~\cite{Iwasaki}:
\begin{figure}[tbh]
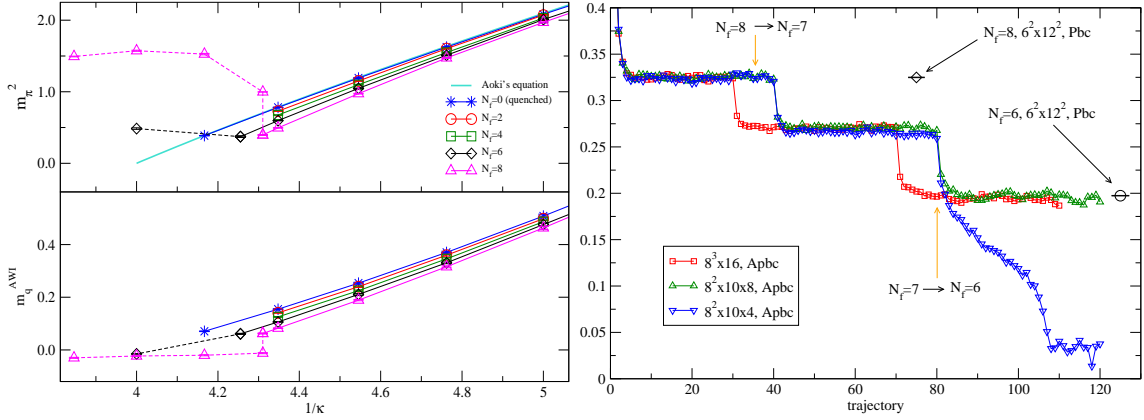
 
\includegraphics[width=7.5cm,height=5.5cm,trim=0 0 0 0,clip=true]{plots/mqmp2-nc3b0_612.eps}
\includegraphics[width=7.5cm,trim=0 0 0 0,clip=true]{plots/fig-plaq-hist.eps}
\caption{$SU(3)$ gauge theory with $N_f$ flavours at $\beta=0$.
{\bf Left panel:} $m_\pi^2$ and $m_q^{\rm AWI}$ as functions of $1/\kappa$.
The analytic prediction 
is shown as a solid blue curve
reaching $m_\pi^2=0$ at $1/\kappa=4$.
{\bf Right panel:} Plaquette history 
in HMC and R-algorithm
in various lattice setups.
}
\label{fig:nc3}
\end{figure}
we did attain thermalization with $N_{CG} < 10000$ in MD step
at $\kappa=0.25$ for $N_f=6$,
and our pion mass is not zero
as shown in the left panel of Fig.~\ref{fig:nc3}.
In addition, our pion and quark masses depend on $N_f$ such that only the
quenched case ($N_f=0$) obeys
Eq.~(\ref{eq:pion}) leading to the conventional Aoki phase.

To understand the origin of this discrepancy 
between our data and theirs,
we monitor the history of the plaquette value
on various lattices with various setups.
In addition to our own HMC simulations with periodic boundary conditions
reported in Ref.~\cite{Nagai}, we now 
monitor the plaquette value
by using MILC code\cite{milc}.
The MILC code uses
the R-algorithm with anti-periodic boundary conditions, which are the same
algorithm and boundaries as were used by Iwasaki {\it et al.} in Ref.~\cite{Iwasaki}.
We generate the configurations
with a cold start for $N_f=8$,
then switch to $N_f=7$ from the last configuration of $N_f=8$
and switch to $N_f=6$ from that of $N_f=7$,
and we plot the plaquette value history
in the right panel of Fig.~\ref{fig:nc3}.
For larger $N_t$,
the simulation is not affected
by the different algorithms and boundary conditions.
However,
we found that 
the plaquette value at $N_t=4$, 
where Iwasaki {\it et al.} obtained their conclusion,
shows a large deviation.

Our observations at $N_t=4$ are consistent 
with the results of Ref.~\cite{Iwasaki},
{\it i.e.} the small value of the plaquette and the pion mass.
However,
it seems clear that
results with this small time extent is not representative of larger lattices;
there seems to be some kind of the thermal phase transition.
We conclude that our studies with larger $N_t$, which showed a massive pion,
are more appropriate for zero-temperature discussions, while conclusions
from Ref.~\cite{Iwasaki} about the conformal window cannot be supported.

\section{$N_c=2$ case: Exploration of the phase structure}
\label{sec:su2}

\begin{figure}[tbh]
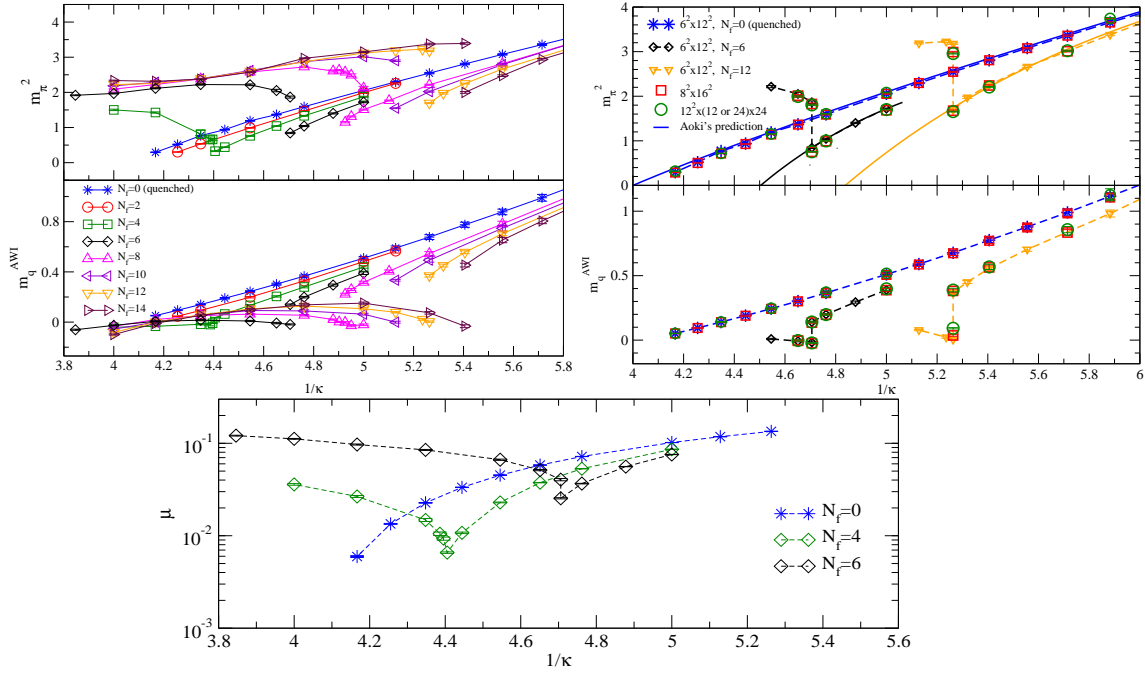

\includegraphics[width=7.5cm,trim=0 0 0 0,clip=true]{plots/mqmp2-nc2b0_612_zoom.eps}
\includegraphics[width=7.5cm,trim=0 0 0 0,clip=true]{plots/mqmp2-nc2b0_latsize.eps} \\
\hspace*{2cm}\includegraphics[width=10cm,height=3.6cm,trim=0 0 0 0,clip=true]{plots/ev-data-su2.eps}
\caption{
$SU(2)$ gauge theory with $N_f$ flavours 
at $\beta=0$ on a $6^2\times12^2$ lattice.
{\bf Upper left panel:} $m_\pi^2$ and $m_q^{\rm AWI}$ as
functions of $1/\kappa$.
{\bf Upper right panel:} 
Volume independence for $N_f=0, 6$  and 12.
The solid blue curve represents 
the prediction 
of the equation.
{\bf Bottom panel:} 
$\mu=\sqrt{\lambda_0(H_W^2)}$,
where $\lambda_0(H_W^2)$ is the lowest eigenvalue of $H_W^2$, 
as a function of $1/\kappa$.
}
\label{fig:nc2}
\end{figure}

For $SU(2)$ gauge theory, Ref.~\cite{Iwasaki} predicted
that $N_f=2$ is in the confinement region
and that $N_f = 3$ is in the conformal window
for the same reason as discussed for $SU(3)$:
$N_{CG} > 10000$.
However, as we have already pointed out in the previous section,
finding $N_{CG}>10000$ during the thermalizing process 
is not reliable evidence for discovering a conformal window.
Although we are unable to obtain direct results for $N_f=2$
due to the very small value of the pion mass 
around $\kappa=0.25$,
we did obtain results for $N_f=4, 6, 8, \cdots$
and we plot those in the upper left panel of Fig.~\ref{fig:nc2}
for various $N_f$ flavours
and in the upper right panel of Fig.~\ref{fig:nc2}
for $N_f=0, 6$ and 12 especially
to check finite size effects.
In the upper panels of Fig.~\ref{fig:nc2},
there are a confinement phase and an unknown massive pion phase,
and these two phases are separated by a transition at
$\kappa=\kappa_d$\footnote{This
$\kappa_d$ is not a precise value.  Hysteresis is observed within a range of
$\kappa$ for $N_f \ge 6$ in $SU(2)$, and $\kappa_d$ is a label for the
location of that transition.}.

Does an Aoki phase exist in the strong coupling region?
The $SU(2)$ case is qualitatively the same as the $SU(3)$ case:
the pion and the quark mass reveal the 2-state signal,
the quenched case ($N_f=0$) agrees with Eq.~(\ref{eq:pion})
but the dynamical case ($N_f >0$) depends on $N_f$,
and further there is no $\kappa_c$ beyond which $m_\pi^2=0$.
Namely our result  is 
\begin{equation}
0.25 = \kappa_c(N_f=0) >  \dots >  \kappa_c(N_f=6) > \dots 
> \kappa_c(N_f=12) >  \dots \,,
\end{equation}
{\it {\bf if there exists $\kappa_c$}} in $N_f > 0$ case.
Therefore it seems there is no standard Aoki phase 
due to non-existence of the massless pion
over the whole region of $\kappa$.

Further we monitor the lowest eigenvalue 
in the bottom panel of Fig.~\ref{fig:nc2},
$\mu=\sqrt{\lambda_0(H_W^2)}$,
where $H_W$ is the hermitian Wilson-Dirac operator,
in order to consider the phase structure
of many Wilson fermions.
The lowest eigenvalue for $N_f >0$
is not zero in both the confinement
and the unknown-massive-pion phase,
while the eigenvalue in $N_f =0$ tends to go to zero 
as $\kappa \rightarrow 0.25$.
According to the modified Banks-Casher relation\cite{cppacs},
the order parameter of the Aoki phase is
$\langle \bar{\psi}(x) \gamma_5 \tau_3 \psi(x) \rangle =
2 \pi \rho_{H_W}(\lambda=0)$
where $\rho_{H_W}(\lambda)$ is the eigenvalue distribution function
of $H_W$.
The data from small lattices suggest that
the order parameter of the Aoki phase is zero,
and this is consistent with the absence of a standard Aoki phase 
in the case with many Wilson fermions.
However,
it would be necessary to investigate on the larger lattices
and to calculate the spectrum of low-lying eigenvalues
before making a strong conclusion, and this is left to future studies.

\begin{figure}[tbh] 
\hspace{2cm}\includegraphics[width=10cm,height=6.7cm,trim=0 0 0 0,clip=true]{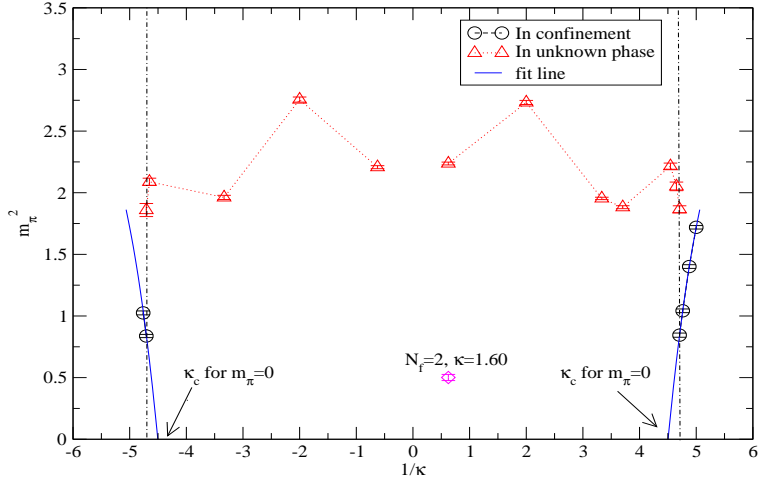}
\caption{
$m_\pi^2$ as a function of not only positive 
but negative $\kappa$ for $N_f=6$ at $\beta=0$
on $6^2 \times12^2$.
}
\label{fig:phase}
\end{figure}
The outline of the phase structure in our research 
is shown in Fig.~\ref{fig:phase}.
The existence of a massive pion phase (the red triangles in the figure)
is apparent, but without that information one might try
to obtain $\kappa_c$ by extrapolating to  $m_\pi^2=0$.
In that way one would find
that there is the gap of $\kappa_c$ ($\kappa_c^{+} \ne \kappa_c^{-}$)
for $m_\pi^2 =0$
between the positive and the negative region.
According to the traditional interpretation,
this is the Aoki phase containing a massless charged pion.
However, the massive pion phase (red triangles) indicates that,
at least with the many Wilson fermions ($N_f/N_c \gg 1$),
another phase has appeared instead of a standard Aoki phase.

Therefore,
in Ref.~\cite{Nagai} and this proceedings,
we don't have a clear signal for the existence of an Aoki phase
and the Sharpe--Singleton first order scenario\cite{sharpe},
but instead we have obtained evidence for a different phase
at strong coupling.

\section{Summary and Discussion}
\label{sec:summary}

We explored $SU(2)$ and $SU(3)$ gauge theories with many Wilson fermions
and reported result in Ref.~\cite{Nagai} and additional results in these
proceedings.
In contrast to Ref.~\cite{Iwasaki}, we find that
(i) The pion mass in the confinement region depends on $N_f$.
(ii) As defined by extrapolation to $m_\pi^2=0$, 
$\kappa_c$ depends on $N_f$ and shifts from 0.25.
(iii) No massless pion exists for any value of $\kappa$. 
Thus the extrapolated $\kappa_c$ is not a very useful quantity, and 
extrapolation to $m_\pi=0$ is not valid when searching for an Aoki phase.
(iv) There is the 2-state signal and hysteresis around $\kappa=\kappa_d$,
which depends on $N_f$.
(v) The $N_f=6$ case in $SU(3)$ gauge theory with Wilson fermions is not 
in the confinement region 
because the pion mass at $\kappa_c=0.25$ is not zero.
Simulations at $N_t=4$ are not representative of larger $N_t$ values, 
so we do not validate the claim\cite{Iwasaki} that 
the critical flavour of the conformal window is 7 ($N_f^{*} \ne 7$).
(vi) There is a standard Aoki phase in the quenched case ($N_f=0$), 
but the situation is markedly different 
in the large flavour case ($N_f/N_c \gg 1$).

Now
we make an effort to identify the unknown massive pion phase.
\begin{figure}[tbh]
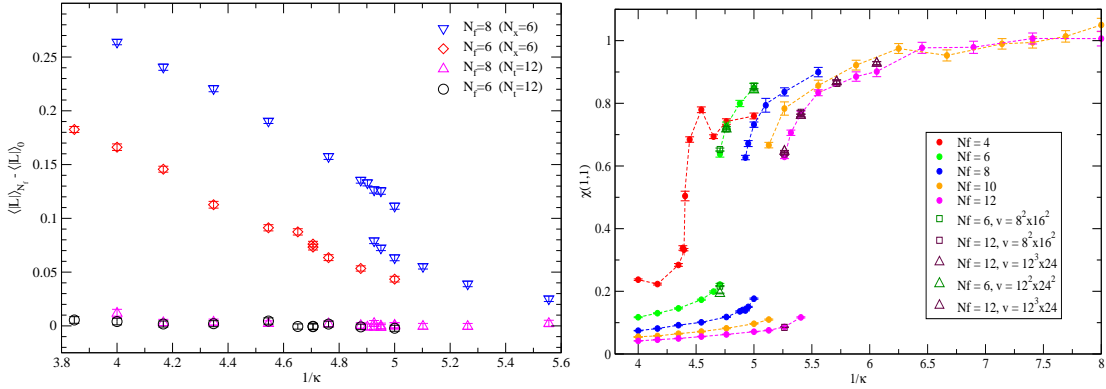
 
\includegraphics[width=7.5cm,trim=0 0 0 0,clip=true]{plots/polyakov.eps}
\includegraphics[width=7cm,height=5cm,trim=0 0 0 0,clip=true]{plots/STZsmr.eps}
\caption{
{\bf Left panel:} Polyakov loop normalized by the quenched case, 
$\langle | L | \rangle_{N_f} - \langle |L| \rangle_{N_f=0}$,
as a function of $1/\kappa$.
Here we take the absolute value of the Polyakov loop.
{\bf Right panel:} Creutz ratio, $\chi(1,1)$, with 10 APE smearings
as a function of $1/\kappa$.
}
\label{fig:add}
\end{figure}
The Polyakov loop is plotted in the left panel of Fig.~\ref{fig:add}.
Along the small extent ($N_x=6$ on $6^2 \times 12^2$),
the Polyakov loop shows finite temperature behaviour
with a 2-state signal and hysteresis.
However, the Polyakov loop along the large extent 
($N_t=12$ on $6^2 \times 12^2$)
doesn't show any transitions
and is consistent with the quenched case ($N_f=0$).
This means
there is no finite temperature transition
on large lattices, so the unknown phase is not the result of a
confinement-deconfinement phase transition.

The right panel of Fig.~\ref{fig:add} is a Creutz ratio, $\chi(1,1)$.
At $\beta=0$ the static quark is confined 
(modulo string breaking),
so the Creutz ratio at short distances
is near unity in lattice units.
We compute $\chi(1,1)$ for $\kappa > \kappa_d$,
and find that it becomes small
in contrast with its value in the confinement phase.
This relative difference might imply
a transition from confinement to weak coupling.

Although we monitored various quantities
to identify the phase,
unfortunately at present
we cannot determine the true nature of this unknown massive-pion phase.
Still,
this phase
is the confirmation 
of a (first order) bulk phase transition
at the strong coupling,
as pointed out in Refs.~\cite{jlqcd,nchrist}
by observing the high- and the low-plaquette phase
and the chiral condensate
with the first order transition.
Our result is the first observation
of that longer list of quantities 
(the pion mass, the quark mass and so on)
and the fact that
the standard Aoki phase seems to be invisible
due to this bulk phase transition
at strong coupling.
Furthermore,
it is found that
this bulk phase transition
even appears
in the strong coupling limit ($\beta=0$)
with Wilson fermions.
This implies
that the (first order) bulk phase transition
is caused by fermionic dynamics only,
not by the gauge dynamics, 
namely this phase is independent of the type of lattice gauge actions.

Although it's difficult to determine which phase is realized,
it will be possible to determine 
whether the standard Aoki phase has appeared or not,
by observing the order parameter of the Aoki phase,
$\langle \bar{\psi} \gamma_5 \tau_3 \psi \rangle$.
Refs.~\cite{Ilgenfritz:2009ns,aokiphase} 
carried out a calculation of the order parameter for $N_f=2$ in $SU(3)$
($N_f/N_c < 1$)
by using the twisted mass Wilson fermions.
Let us remember again the bottom panel of Fig.~\ref{fig:nc2}.
When this figure is turned upside-down, 
a result similar to the data of the order parameter
might be obtained in $SU(2)$,
due to the very rough relation $\langle \bar{\psi} \gamma_5 \tau_3 \psi \rangle \sim \frac{1}{\mu}$ on finite lattices.
If this is true, $SU(2)$ twisted mass fermions might show a bulk phase transition
in $N_f=2$, instead of Aoki phase.
Therefore we are calculating the order parameter
in $N_f=2$ and $N_f \ge 6$ in $SU(2)$ gauge theories ($N_f/N_c \ge 1$)
with twisted mass Wilson fermions.
It should be possible to clarify the phase structure
of a system with (many) Wilson fermions
in the near future.

\section*{Acknowledgments}
\label{sec:acknowledge}
This work was supported in part by the Natural Sciences and Engineering
Research Council of Canada, and was made possible by the facilities of the
Shared Hierarchical Academic Research Network (SHARCNET:www.sharcnet.ca).
We acknowledge use of Regina's VXRACK computer cluster.
A part of this job is done by the version 7 
of the publicly available code of the MILC collaboration\cite{milc}.
G.K.\ thanks the University of Regina for hospitality during her stay there.
K.N.\ thanks Norikazu Yamada 
for the enlightening discussion
of the first order bulk phase transition.
K.N.\ is supported in part 
by the Grant-in-Aid for Scientific Research on Innovative Areas in Japan
(No. 20105002).


\end{document}